\documentclass[twocolumn,prl,superscriptaddress,float,aps]{revtex4-2}
\usepackage{graphicx,amsfonts,amssymb,amsmath,hyperref,hypcap,enumerate}
\usepackage{color}
\usepackage{bm}
\usepackage{multirow}
\usepackage{makecell}
\usepackage{color}
\usepackage{orcidlink}
\usepackage{hyperref}
\usepackage[title]{appendix}
\hypersetup{
colorlinks = true,
linkcolor = [rgb]{0.70,0.13,0.13},
citecolor = [rgb]{0.13,0.55,0.13},
urlcolor = [rgb]{0.25, 0.41, 0.88}}
\usepackage{makecell}
\usepackage{diagbox}

\begin{document}

\title{Composite Boson Theory of Fractional Chern Insulators}
\author{Guangyu Yu\orcidlink{0009-0000-7405-1853}}
\affiliation{Kavli Institute for Theoretical Sciences, University of Chinese Academy of Sciences, Beijing 100190, China}

\author{Zheng Zhu\orcidlink{0000-0001-7510-9949}}
\email{zhuzheng@ucas.ac.cn}
\affiliation{Kavli Institute for Theoretical Sciences, University of Chinese Academy of Sciences, Beijing 100190, China}

\date{\today}

\begin{abstract}

The understanding of fractional Chern insulators (FCIs) has been deeply guided by band topology and quantum geometry. Here, we introduce a real-space theoretical framework in which FCIs are understood in terms of composite bosons, local objects consisting of electrons bound to their energetically excluded surrounding orbitals. The central element of our framework is the construction of a radially ordered set of maximally localized basis for Chern bands without requiring continuous rotational symmetry. Within this basis, the complex many-body problem simplifies to a real-space organizing principle: a stable FCI occurs if the orbitals excluded around central electrons are those maximizing the two-body interaction energy. We validate this with direct numerical evidence for composite boson formation in the Haldane model, demonstrating that our criterion reliably characterizes FCIs. Importantly, our analysis illustrates that the composite boson framework bridges the fractional quantum Hall effect in continuum and lattice paradigms, providing a unified and intuitive real-space interpretation for distinct correlated phases. It thus establishes a foundation for diagnosing and guiding the design of both Abelian and non-Abelian topologically ordered phases across distinct platforms.

\end{abstract}

\maketitle

\emph{Introduction.---} 
Two-dimensional topological phases have attracted intense interest in recent years, partly driven by experimental advances such as the observation of fractional quantum anomalous Hall effect (FQAHE) in moiré heterostructures~\cite{cai2023signatures,zeng2023thermodynamic,park2023observation,PhysRevX.13.031037}. The fractional quantum Hall (FQH) effect~\cite{chakraborty2013quantum,prange1987quantum,jain2007composite}, a cornerstone of topological physics, continues to challenge and enrich our understanding of strongly correlated systems. Fractional Chern insulators (FCIs)~\cite{PhysRevLett.106.236804}, 
lattice analogs of FQH states, provide a platform to explore analogous physics without relying on certain auxiliary elements inherent to Landau levels, such as continuous rotational symmetry~\cite{PhysRevLett.107.116801} and holomorphic wavefunctions~\cite{10.1063/1.5046122}.

Conventional studies of FCIs often prioritize band properties~\cite{PhysRevB.90.165139,PhysRevB.104.045104,PhysRevB.108.205144},
while assuming a ``natural" form of interaction, such as Coulomb or screened Coulomb potentials. 
Despite their success in explaining many observations, these frameworks face several inherent limitations. In particular, they may overlook exotic phases that could emerge from intentionally engineered interactions~\cite{PhysRevLett.132.236503}. 
More fundamentally, since the ground state is determined solely by the matrix elements of projected interactions~\cite{PhysRevLett.51.605}, reliance on embedding-dependent quantities can obscure the essential origin of FCIs~\cite{PhysRevB.102.165148}.

\begin{figure}
    \centering
    \includegraphics[width=1\linewidth]{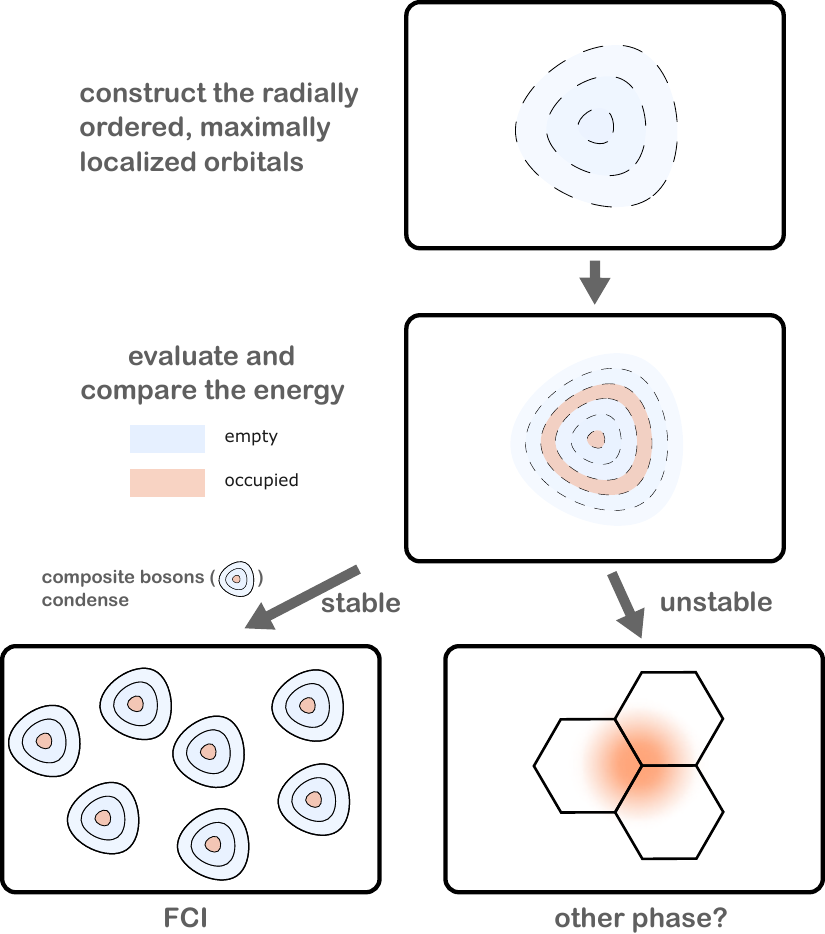}
    \caption{
     {Conceptual workflow of the composite boson theory for fractional Chern insulators (FCIs).} The diagram outlines the key steps of the theoretical framework: (i) Construction of a radially ordered set of maximally localized orbital basis within a topological flat band.
    (ii) Calculation and comparison of the two-body interaction energy between an electron in the central orbital ($m=0$) and a test electron in the $m$-th orbital. A stable composite boson is formed if the orbitals it excludes (e.g., $m=1, 2$ for Laughlin state) are those maximizing the two-body interaction energy. (iii) Condensation of these stable composite bosons gives rise to the  FCI phase. This workflow establishes a universal real-space organizing principle for understanding and predicting FCIs.}
    \label{fig:flowchart}
\end{figure}

Here, we overcome these limitations by developing a real-space theory for FCIs based on the composite boson paradigm~\cite{PhysRevLett.62.82,PhysRevLett.62.86}. In this framework, a composite boson refers to an effective degree of freedom defined by an electron bound to its energetically excluded surrounding orbitals.
The key to this theory is the construction of a radially ordered set of maximally localized orbital basis. We postulate that a stable FCI  corresponds to a condensate of composite bosons, whose stability is governed by a simple energy criterion. The framework is validated numerically using the Haldane model, providing direct evidence of composite boson formation and FCI emergence. We further apply it to explain distinct behaviors across different Landau levels, highlighting a unified organizing principle from a real-space perspective, and discuss broader implications. The conceptual workflow of this theoretical framework is summarized in Fig.~\ref{fig:flowchart}.

\emph{Real-space composite boson framework.---}
We understand the FCIs based on a radially ordered set of maximally localized orbitals in real space, analogous to Landau level wavefunctions in the symmetric gauge. For a lattice system, we begin by constructing an orthogonal basis spanning the target band, typically a topologically flat band. This is implemented by diagonalizing the band projection operator $\hat{P}^{(n)} = \sum_{\mathbf{k} \in \text{BZ}} |\psi_{n, \mathbf{k}}\rangle \langle \psi_{n, \mathbf{k}}|$, the eigenvectors associated with nonzero eigenvalues of $\hat{P}^{(n)}$ constitute an orthogonal basis spanning the band. 
To impart a real-space hierarchy to this basis, we diagonalize a suitable measure of spatial spread within the band subspace. A natural choice is the squared position operator $\hat{r}^2$, or its lattice analog $\sum_i \hat{n}_i (\mathbf{R}_i - \mathbf{R}_0)^2$, which measures the mean squared distance of an orbital from a chosen origin $\mathbf{R}_0$. The eigenvectors of this operator are automatically sorted by their distance from the origin, yielding the desired radially ordered, maximally localized basis.

Within this framework, we propose that the FCI state exhibits a distinctive internal structure: electrons in localized orbitals bind to their surrounding exclusion zones to form composite objects. Taking the $\nu=1/3$ state as a paradigmatic example, occupation of a central orbital necessitates that the subsequent two orbitals remain empty. These unoccupied orbitals are effectively attached to the central electron, forming a composite boson. While similar structural motifs are well-established in the LLL context~\cite{PhysRevLett.61.1985,Johri_2016}, we here generalize this framework to the discrete lattice systems of FCIs. 

The second crucial postulate is that the stability of an FCI state is governed by the robustness of its constituent composite bosons. If their integrity is maintained against fluctuations, they may undergo Bose-Einstein condensation, thereby establishing the many-body FCI ground state. This perspective reduces the complex problem of FCI stability to evaluating the energetic favorability of an individual composite boson. To this end, we consider a two-electron configuration where one electron occupies the central orbital ($m=0$) and a test electron occupies the $m$-th orbital. The interaction energy of this configuration is given by $U_m=\langle 0|c_0 c_m \hat Vc^\dagger_m c_0^\dagger|0\rangle$, where $\hat{V}$ denotes the two-body interaction and $|0\rangle$ the vacuum. 
A composite boson is deemed stable if the orbitals excluded around one or more central electrons are those maximizing the two-body interaction energy.
This energy-based criterion provides a transparent and intuitive foundation for predicting the emergence of FCIs under generic interactions.

\begin{figure*}
    \centering
    \includegraphics[width=\linewidth]{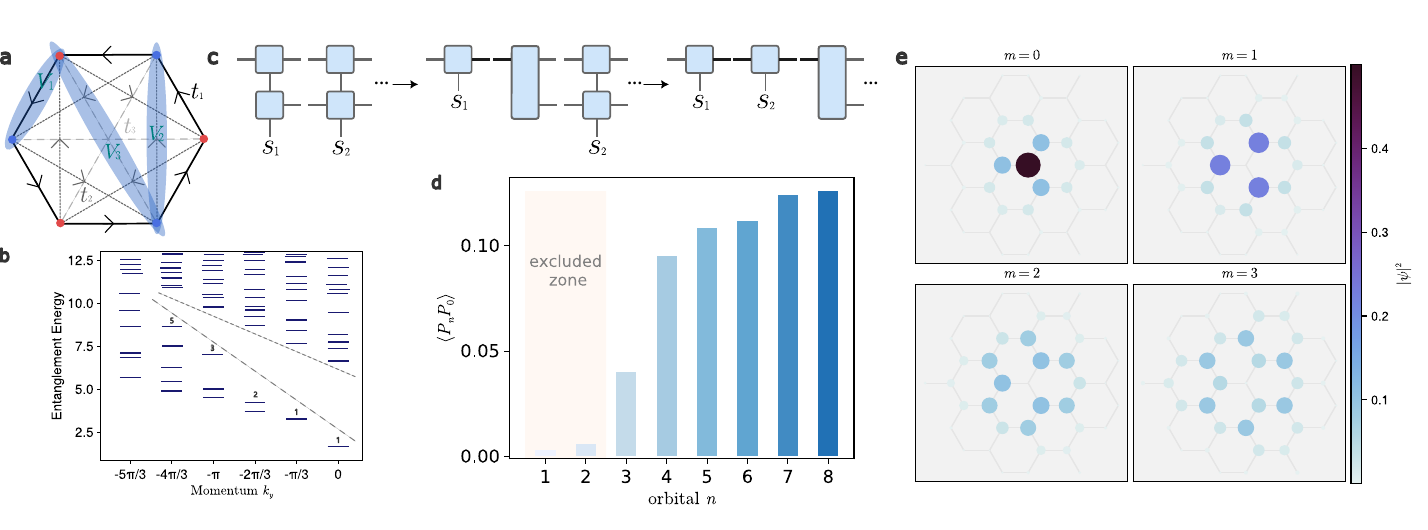}
    \caption{
     {Numerical implementation and evidence for the real-space composite-boson framework.}
    (a) Lattice geometry and parameters of the Haldane model. (b) The entanglement spectrum of the many-body ground state at filling $\nu=1/3$, exhibiting the characteristic low-lying level counting $1, 1, 2, 3, 5, ...$, a hallmark of the underlying topological order and strong evidence for the FCI phase.
    (c) Schematic of the matrix product state (MPS)  compression scheme used to measure the non-local occupation correlation $\langle \Psi|P_n P_0|\Psi\rangle$.
    (d) Measured occupation correlation $\langle \Psi|P_n P_0|\Psi\rangle$, showing a pronounced suppression of electron occupation for the first two orbitals ($n=1,2$) adjacent to an occupied central orbital. This pattern provides direct numerical evidence for real-space composite-boson formation, analogous to flux attachment in the fractional quantum Hall effect. (e) The first few radially ordered, maximally localized orbitals, labeled by the orbital index $m = 0, 1, 2, \ldots$, near the chosen origin.
    }
    \label{fig:2}
\end{figure*}

\emph{Composite boson and FCI in the Haldane model.---} 
We validate the composite boson framework using the Haldane model~\cite{PhysRevLett.61.2015}, a prototypical lattice model hosting  FCIs. The Hamiltonian is 
\begin{equation}
    \begin{aligned}
        H = &\sum_{ij} t_{ij}c^\dagger_i c_j + \mathrm{h.c.}
        +\sum_{ij} V_{ij} \hat n_i \hat n_j,
    \end{aligned}
\end{equation}
with hopping parameters and lattice geometry illustrated in Fig.~\ref{fig:2}(a). To isolate flat band physics dominated by interactions, we choose parameters $t_1=-1,t_2=-0.6\exp(i0.4\pi),t_3=0.58$ that yield a flat band with a flatness ratio exceeding 50 and non-zero Chern numbers, providing an ideal platform for FCIs~\cite{PhysRevLett.107.146803}.

We first construct the radially ordered, maximally localized basis within the flat band on a cylinder of circumference $L_y = 6$, a geometry suited for implementing matrix product state (MPS) techniques to probe the internal structure of FCIs. As outlined in the theoretical framework, we begin by diagonalizing the projection operator $\hat P^{(n)}$ onto the lower band to obtain an initial orthonormal basis. Since the composite boson picture is inherently local, we are primarily interested in orbitals concentrated near the central site. To this end, we employ a real-space localization procedure. We restrict the Hilbert space to a finite circular region. 
For each site $m$ within this region, we consider the state $\hat P^{(n)}|m\rangle$ where $|m\rangle$ denotes the atomic orbital, and discard components outside the circle, yielding the truncated state $|\tilde{m}\rangle$. This truncation primarily affects orbitals near the boundary,
preserving the accuracy of the wavefunctions near the center. 
We then diagonalize the band projector within this restricted subspace. Specifically, we diagonalize the matrix
$S_{mn}=\langle m|\hat P^{(n)}|n\rangle=\langle \tilde{m}|\tilde{n}\rangle$ and $\sum_n S_{mn} w^{(\lambda)}_n=\lambda w^{(\lambda)}_m$. 
The resulting eigenvectors form an orthonormal set of functions $|\psi_\lambda\rangle=\frac{1}{\sqrt{\lambda}}\sum_m w^{(\lambda)}_m|\tilde{m}\rangle$ that are localized around the origin. This procedure is analogous to the construction of Boys orbitals in quantum chemistry~\cite{RevModPhys.32.296}. Finally, we recombine them into the maximally localized basis by diagonalizing the operator $\sum_i \hat n_i (R_i-R_0)^2$ within this set. The first few orbitals are shown in Fig.~\ref{fig:2}(e), illustrating the desired radial ordering.

Equipped with this spatially resolved basis, we probe the internal structure of the FCI state. We compute the occupation correlation $\langle\Psi|P_nP_0|\Psi\rangle$, where $|\Psi\rangle$ is the ground state of FCI, $P_m=c^\dagger_mc_m$ projects onto the $m$-th ($m\ge0$) orbitals and $c_m^\dagger$ is the corresponding creation operator. This correlation measures the likelihood of finding an electron in orbital $n$ given that the central orbital ($m=0$) is occupied, directly revealing the exclusion pattern around a central electron. 
In practice, the measurement of such non-local observables poses significant computational challenges. Although the projectors can in principle be expressed as matrix product operators (MPOs), the bond dimension of the resulting MPO increases rapidly with the size of the support region. To address this issue, we develop a compression strategy: each projector is represented as an MPO and applied sequentially. After each application, we perform a ``zip-up"~\cite{Stoudenmire_2010} style compression on the resulting MPS
to control the bond dimension, as illustrated in Fig.~\ref{fig:2}(c) using the first two sites as an example. Specifically, we first apply an MPO to an MPS and perform a singular value decomposition (SVD) to obtain an MPS where one leg carries a high bond dimension (represented by a bold line). We then contract the remainder of the decomposition with the MPS and MPO at the next site and perform another SVD, yielding an MPS with two legs of elevated yet manageable bond dimensions. This procedure is repeated until all MPOs have been applied. The iterative compression scheme prevents the bond dimensions from growing prohibitively, rendering the computation feasible. In our calculations, the ground-state bond dimensions remain moderate, allowing lossless compression throughout.

We tune the filling to $\nu = 1/3$ with $V_1=1, V_2=V_3=0$, and take the resulting Laughlin state, which is identified by the entanglement spectrum~\cite{PhysRevLett.101.010504} in Fig.~\ref{fig:2}(b), as a representative example.
The direct measurement of the occupation correlation $\langle\Psi|P_nP_0|\Psi\rangle$ yields a definitive signature of composite boson formation.
Fig.~\ref{fig:2}(d) displays a pronounced suppression of occupation precisely for the first two orbitals ($n=1,2$) adjacent to the occupied central orbital.
This indicates that once a central orbital is occupied, its immediate neighboring orbitals are effectively excluded from being occupied by other electrons. These empty orbitals are thus bound to the central electron, forming a composite object. This real-space binding mechanism is the direct lattice analog of flux attachment in the FQH effect. We therefore identify this bound entity formed by an electron and its associated exclusion zone as a composite boson.

\emph{Unified composite boson framework for continuum and lattice FQH states.---} 
In the context of Landau levels (LLs), the Laughlin state is commonly understood as the zero-energy ground state of a specific Haldane pseudopotential~\cite{PhysRevLett.51.605}. Realistic interactions are then treated as perturbations to this model Hamiltonian, and phases adiabatically connected to the Laughlin state are identified as belonging to the same topological class. This framework crucially relies on the continuous rotational symmetry of the LLs, which enables a classification of pseudopotentials by relative angular momentum. 
In lattice-based FCIs, however, continuous rotational symmetry is generally absent, rendering such a classification scheme ambiguous. Although generalized pseudopotentials can be constructed for any Chern bands~\cite{PhysRevLett.107.116801,PhysRevLett.118.146403}, the resulting descriptions lack the intuitive physical clarity compared with the continuum LL setting.

To elucidate the composite bosons as a unified real-space organizing principle, we first analyze the continuum Landau-level FQH states and then return to the lattice FCI counterparts.

\begin{table}[htbp]
\centering
\begin{tabular}{|c|c|c|c|c|}
\hline
\diagbox[width=1.0cm, height=1.5em]{n}{m} & 1 & 2 & 3 & 4 \\
\hline
1 & 0.44311 & 0.44310 & 0.40156 & 0.36002 \\
\hline
2 & 0.34011 & 0.36521 & 0.34559 & 0.32378 \\
\hline
3 & 0.28601 & 0.31194 & 0.29722 & 0.28322 \\
\hline
\end{tabular}
\caption{The effective interaction energy $U_m^{(n)}$ between the central orbital (m=0) and orbital $m$ in $n$-th LL. $U_m^{(n)}$ are given in units of $e^2/{\epsilon l_B}$. Data show the energy values used to compare stability patterns across levels, e.g., decreasing $U_m^{(1)}$ in LLL supports composite boson formation, while variations in higher levels explain phase differences.}
\label{table:LL}
\end{table}

In the lowest Landau level (LLL), the most compact orbitals coincide with the symmetric-gauge Landau-level wavefunctions, which are naturally ordered from the innermost to the outermost.
The spatial area of each orbital is approximately one flux quantum. Once the central orbital is occupied, we evaluate the two-body interaction associated with the other particle occupying orbital $m$, $U_m^{(n)}=\langle 0m|\hat{V}|0m\rangle$, where $\hat V=1/|\hat{r}_1-\hat{r}_2|$ denotes the Coulomb interaction, and $|0m\rangle$ represents antisymmetrized two-particle states with orbital $0$ and $m$ occupied.
As shown in Table.~\ref{table:LL}, $U_m^{(1)}$ decreases with increasing orbital index $m$. This implies that, for $\nu=1/3$, excluding the two nearest neighboring orbitals minimizes the interaction energy, thereby stabilizing a composite boson formed by an electron and three successive orbitals. Condensation of these composite bosons gives rise to the Laughlin state. Chern bands that emulate the LLL structure thus naturally favor the same state under Coulomb interactions, with the essential correspondence being the monotonic decrease of $U_m^{(1)}$ with $m$.

The situation differs significantly in the second Landau level (2LL), where the central orbital is less localized and the subsequent orbitals are effectively pushed outward, leading to a distinct pattern of $U_m^{(2)}$. The $z$ components of angular momentum for orbitals $m=0,1,2,\dots$ are $L_z=0,\pm1,2,\dots$, and notably $U_2^{(2)}>U_1^{(2)}$.
This provides an intuitive real-space explanation for the difference
between the LLL and the 2LL. Under Coulomb interaction, a half-filled LLL forms a composite fermion liquid (CFL), whereas the 2LL favors the Moore-Read (MR) state~\cite{MOORE1991362}. 
For the MR state,
attaching the $m=1$ orbital to the central electron is energetically unfavorable, as occupying $m=2$ incurs a higher energy cost.  
Electrons thus occupy $m=0,1$ and exclude the occupation of
$m=2$, forming a composite boson consisting of two electrons bound to four orbitals.
Although $U_3^{(2)}>U_1^{(2)}$, the stability criterion is still satisfied. This is because when constructing a two-electron composite boson, interactions between the second (central) electron and the excluded orbitals must be included. Our calculations reveal that the $L_z=-1$ orbital interacts more strongly with orbitals $m=2,3,\dots$ than $L_z=1$. This lifts the degeneracy of $U_1^{(2)}$ and confirms that the composite boson is formed by electrons in $L_z=-1,0$, while the $L_z=1,2$ orbitals, which maximize the total interaction energy, are excluded.
In contrast, in the LLL, attaching two orbitals to a single electron does not produce a boson but remains energetically favorable, stabilizing the CFL. Unlike pseudopotential analyses, which show $V_1 > V_3$ in both LLs~\cite{PhysRevB.78.155308}, our real-space framework captures the essential distinction and guides the identification of MR-like states.

In the third Landau level (3LL), the maximum of $U_m^{(3)}$ occurs at $m=2$, corresponding to the fourth and fifth orbitals, given that $L_z=0,\pm1,\pm2,3,\cdots$ for $n=2$. At $\nu=1/3$, forming a composite boson comprising one electron and three successive orbitals is energetically unfavorable, and for $\nu=1/2$, a composite boson of two electrons and four successive orbitals is similarly disfavored. As a result, in the 3LL the composite boson is unstable, and the system gives way to the charge-density-wave (CDW) phases~\cite{PhysRevLett.82.394,PhysRevLett.83.1219,PhysRevLett.85.5396,PhysRevB.104.L121110,yu2025deconfinedquantumcriticalpoint}. 
Also note that in 2LL at filling $\nu=1/3$, the Coulomb interaction does not explicitly favor the formation of stable single-electron composite bosons, whereas two-electron composite bosons remain possible. We remark that the topological nature of the corresponding FCI is still an open question~\cite{PhysRevB.92.035103,PhysRevB.105.165145}, which ultimately requires an investigation of the statistics of its quasiparticle excitations, a direction we leave for future work.

\begin{table}[htbp]
\centering
\begin{tabular}{|c|c|c|c|c|c|c|} 
\hline 
m & 1 & 2 & 3 & 4 & 5 & 6\\
\hline
$U_m$ & 0.37720 & 0.24609 & 0.18980 & 0.08310 & 0.06785 & 0.05737\\
\hline
\end{tabular}
\caption{Effective interaction energy $U_m$ between the central orbital ($m=0$) and the $m$-th orbital in the FCI (Haldane model). Values indicate that orbitals $m=1$ and 2 have the highest energy, favoring their exclusion to form composite bosons, analogous to the Laughlin state in Landau levels.}
\label{table:FCI}
\end{table}

In FCIs, additional factors require careful consideration. Finite band dispersion can blur the composite-boson picture, so we focus on nearly flat bands where interactions dominate. In our example Haldane model, we evaluate the two-body interaction $U_m=\langle0m|\hat{V}|0m\rangle$ with $\hat{V}=\sum_{ij}V_{ij} \hat n_{i}\hat n_{j}$ and $V_1=1$. As shown in Table.~\ref{table:FCI}, $U_m$ decreases with increasing  $m$. At $\nu = 1/3$, the nearest-neighbor orbitals $m=1,2$ are therefore energetically excluded, giving rise to a composite boson comprising the central electron and three successive orbitals.

Given the numerical evidence above, one might expect stabilizing the non-Abelian MR state using short-range two-body interactions in the Haldane model. However, with interaction up to $V_3$, the system always favors CDWs in the region where composite bosons are energetically preferred. At $\nu = 1/2$, two-orbital models admit multiple CDW configurations, which generally have lower energy than the MR state. 
We note that the MR state was recently realized in a flattened Kagome-lattice band with short range two-body interactions~\cite{fonseca2025gradientbasedsearchquantumphases}. In that setup, the many-body wavefunction closely resembles an LLL-like state, and the combination of a large $V_3$ and small $V_2$ raises the energy of the third and fourth orbitals above the second. Within our composite-boson framework, these seemingly unnatural interactions could be interpreted as stabilizing the required composite-boson structure for the MR state.

\emph{Summary and Outlook.---}
We have established a unified and efficient real-space framework that elucidates FCIs through the condensation of composite bosons, the fundamental building blocks of FCIs in this theory. The core of our approach lies in constructing a radially ordered, maximally localized basis, which enables direct probing of composite boson formation in microscopic lattice models using MPS techniques. The composite boson theory bridges the continuum Landau levels and lattice systems. It not only explains the differences across various Landau levels by revealing distinct energy patterns, but also provides a predictive tool for FCIs. 

This real-space composite boson framework opens several compelling avenues. Its computational efficiency, which hinges on simple two-particle calculations, positions it as an ideal tool for high-throughput exploration of topological phases in a vast landscape of materials, from twisted moiré systems to engineered lattices. 
Another particularly promising application lies in the targeted design of interactions to stabilize exotic states, which could be realized in cold-atom platforms. It opens the door to the reverse engineering of interactions to stabilize desired non-Abelian states. Moreover, its conceptual foundation paves the way for future work in more complex scenarios, including wavefunction construction~\cite{He_2015} and systems with higher Chern numbers or multi-component quantum Hall systems, potentially uncovering new phases of matter beyond the conventional FCIs and enriching our understanding of topological order.

\textit{Acknowledgement.---}
This work was supported by the National Natural Science Foundation of China (Grant No. 92477106) and the Fundamental Research Funds for the Central Universities.

\bibliography{ref} 

@article{PhysRevLett.51.605,
  title = {Fractional Quantization of the Hall Effect: A Hierarchy of Incompressible Quantum Fluid States},
  author = {Haldane, F. D. M.},
  journal = {Phys. Rev. Lett.},
  volume = {51},
  issue = {7},
  pages = {605--608},
  numpages = {0},
  year = {1983},
  month = {Aug},
  publisher = {American Physical Society},
  doi = {10.1103/PhysRevLett.51.605},
  url = {https://link.aps.org/doi/10.1103/PhysRevLett.51.605}
}

@article{PhysRevLett.61.1985,
  title = {Off-Diagonal Long-Range Order in Fractional Quantum-Hall-Effect States},
  author = {Rezayi, E. H. and Haldane, F. D. M.},
  journal = {Phys. Rev. Lett.},
  volume = {61},
  issue = {17},
  pages = {1985--1988},
  numpages = {0},
  year = {1988},
  month = {Oct},
  publisher = {American Physical Society},
  doi = {10.1103/PhysRevLett.61.1985},
  url = {https://link.aps.org/doi/10.1103/PhysRevLett.61.1985}
}

@article{Johri_2016,
doi = {10.1088/1367-2630/18/2/025011},
url = {https://doi.org/10.1088/1367-2630/18/2/025011},
year = {2016},
month = {feb},
publisher = {IOP Publishing},
volume = {18},
number = {2},
pages = {025011},
author = {Johri, Sonika and Papić, Z and Schmitteckert, P and Bhatt, R N and Haldane, F D M},
title = {Probing the geometry of the Laughlin state},
journal = {New Journal of Physics}
}

@article{PhysRevLett.61.2015,
  title = {Model for a Quantum Hall Effect without Landau Levels: Condensed-Matter Realization of the "Parity Anomaly"},
  author = {Haldane, F. D. M.},
  journal = {Phys. Rev. Lett.},
  volume = {61},
  issue = {18},
  pages = {2015--2018},
  numpages = {0},
  year = {1988},
  month = {Oct},
  publisher = {American Physical Society},
  doi = {10.1103/PhysRevLett.61.2015},
  url = {https://link.aps.org/doi/10.1103/PhysRevLett.61.2015}
}

@article{PhysRevLett.107.146803,
  title = {Fractional Quantum Hall Effect of Hard-Core Bosons in Topological Flat Bands},
  author = {Wang, Yi-Fei and Gu, Zheng-Cheng and Gong, Chang-De and Sheng, D. N.},
  journal = {Phys. Rev. Lett.},
  volume = {107},
  issue = {14},
  pages = {146803},
  numpages = {5},
  year = {2011},
  month = {Sep},
  publisher = {American Physical Society},
  doi = {10.1103/PhysRevLett.107.146803},
  url = {https://link.aps.org/doi/10.1103/PhysRevLett.107.146803}
}

@article{RevModPhys.32.296,
  title = {Construction of Some Molecular Orbitals to Be Approximately Invariant for Changes from One Molecule to Another},
  author = {Boys, S. F.},
  journal = {Rev. Mod. Phys.},
  volume = {32},
  issue = {2},
  pages = {296--299},
  numpages = {0},
  year = {1960},
  month = {Apr},
  publisher = {American Physical Society},
  doi = {10.1103/RevModPhys.32.296},
  url = {https://link.aps.org/doi/10.1103/RevModPhys.32.296}
}

@article{Stoudenmire_2010,
doi = {10.1088/1367-2630/12/5/055026},
url = {https://doi.org/10.1088/1367-2630/12/5/055026},
year = {2010},
month = {may},
publisher = {},
volume = {12},
number = {5},
pages = {055026},
author = {Stoudenmire, E M and White, Steven R},
title = {Minimally entangled typical thermal state algorithms},
journal = {New Journal of Physics}
}

@article{PhysRevLett.118.146403,
  title = {Generalized Pseudopotentials for the Anisotropic Fractional Quantum Hall Effect},
  author = {Yang, Bo and Hu, Zi-Xiang and Lee, Ching Hua and Papi\ifmmode \acute{c}\else \'{c}\fi{}, Z.},
  journal = {Phys. Rev. Lett.},
  volume = {118},
  issue = {14},
  pages = {146403},
  numpages = {5},
  year = {2017},
  month = {Apr},
  publisher = {American Physical Society},
  doi = {10.1103/PhysRevLett.118.146403},
  url = {https://link.aps.org/doi/10.1103/PhysRevLett.118.146403}
}

@article{PhysRevLett.101.010504,
  title = {Entanglement Spectrum as a Generalization of Entanglement Entropy: Identification of Topological Order in Non-Abelian Fractional Quantum Hall Effect States},
  author = {Li, Hui and Haldane, F. D. M.},
  journal = {Phys. Rev. Lett.},
  volume = {101},
  issue = {1},
  pages = {010504},
  numpages = {4},
  year = {2008},
  month = {Jul},
  publisher = {American Physical Society},
  doi = {10.1103/PhysRevLett.101.010504},
  url = {https://link.aps.org/doi/10.1103/PhysRevLett.101.010504}
}

@article{PhysRevLett.107.116801,
  title = {Geometrical Description of the Fractional Quantum Hall Effect},
  author = {Haldane, F. D. M.},
  journal = {Phys. Rev. Lett.},
  volume = {107},
  issue = {11},
  pages = {116801},
  numpages = {5},
  year = {2011},
  month = {Sep},
  publisher = {American Physical Society},
  doi = {10.1103/PhysRevLett.107.116801},
  url = {https://link.aps.org/doi/10.1103/PhysRevLett.107.116801}
}

@article{PhysRevLett.82.394,
  title = {Evidence for an Anisotropic State of Two-Dimensional Electrons in High Landau Levels},
  author = {Lilly, M. P. and Cooper, K. B. and Eisenstein, J. P. and Pfeiffer, L. N. and West, K. W.},
  journal = {Phys. Rev. Lett.},
  volume = {82},
  issue = {2},
  pages = {394--397},
  numpages = {0},
  year = {1999},
  month = {Jan},
  publisher = {American Physical Society},
  doi = {10.1103/PhysRevLett.82.394},
  url = {https://link.aps.org/doi/10.1103/PhysRevLett.82.394}
}

@article{PhysRevLett.83.1219,
  title = {Charge-Density-Wave Ordering in Half-Filled High Landau Levels},
  author = {Rezayi, E. H. and Haldane, F. D. M. and Yang, Kun},
  journal = {Phys. Rev. Lett.},
  volume = {83},
  issue = {6},
  pages = {1219--1222},
  numpages = {0},
  year = {1999},
  month = {Aug},
  publisher = {American Physical Society},
  doi = {10.1103/PhysRevLett.83.1219},
  url = {https://link.aps.org/doi/10.1103/PhysRevLett.83.1219}
}

@article{MOORE1991362,
title = {Nonabelions in the fractional quantum hall effect},
journal = {Nuclear Physics B},
volume = {360},
number = {2},
pages = {362-396},
year = {1991},
issn = {0550-3213},
doi = {https://doi.org/10.1016/0550-3213(91)90407-O},
url = {https://www.sciencedirect.com/science/article/pii/055032139190407O},
author = {Gregory Moore and Nicholas Read}
}

@misc{fonseca2025gradientbasedsearchquantumphases,
      title={Gradient-based search of quantum phases: discovering unconventional fractional Chern insulators}, 
      author={André Grossi Fonseca and Eric Wang and Sachin Vaidya and Patrick J. Ledwith and Ashvin Vishwanath and Marin Soljačić},
      year={2025},
      eprint={2509.10438},
      archivePrefix={arXiv},
      primaryClass={cond-mat.str-el},
      url={https://arxiv.org/abs/2509.10438}, 
}

@article{PhysRevB.78.155308,
  title = {Orbital Landau level dependence of the fractional quantum Hall effect in quasi-two-dimensional electron layers: Finite-thickness effects},
  author = {Peterson, Michael R. and Jolicoeur, Th. and Das Sarma, S.},
  journal = {Phys. Rev. B},
  volume = {78},
  issue = {15},
  pages = {155308},
  numpages = {25},
  year = {2008},
  month = {Oct},
  publisher = {American Physical Society},
  doi = {10.1103/PhysRevB.78.155308},
  url = {https://link.aps.org/doi/10.1103/PhysRevB.78.155308}
}

@article{PhysRevB.105.165145,
  title = {Abelian origin of $\ensuremath{\nu}=2/3$ and $2+2/3$ fractional quantum Hall effect},
  author = {Hu, Liangdong and Zhu, W.},
  journal = {Phys. Rev. B},
  volume = {105},
  issue = {16},
  pages = {165145},
  numpages = {20},
  year = {2022},
  month = {Apr},
  publisher = {American Physical Society},
  doi = {10.1103/PhysRevB.105.165145},
  url = {https://link.aps.org/doi/10.1103/PhysRevB.105.165145}
}

@article{PhysRevB.90.165139,
  title = {Band geometry of fractional topological insulators},
  author = {Roy, Rahul},
  journal = {Phys. Rev. B},
  volume = {90},
  issue = {16},
  pages = {165139},
  numpages = {7},
  year = {2014},
  month = {Oct},
  publisher = {American Physical Society},
  doi = {10.1103/PhysRevB.90.165139},
  url = {https://link.aps.org/doi/10.1103/PhysRevB.90.165139}
}

@article{PhysRevB.104.045104,
  title = {K\"ahler geometry and Chern insulators: Relations between topology and the quantum metric},
  author = {Mera, Bruno and Ozawa, Tomoki},
  journal = {Phys. Rev. B},
  volume = {104},
  issue = {4},
  pages = {045104},
  numpages = {13},
  year = {2021},
  month = {Jul},
  publisher = {American Physical Society},
  doi = {10.1103/PhysRevB.104.045104},
  url = {https://link.aps.org/doi/10.1103/PhysRevB.104.045104}
}

@article{PhysRevB.108.205144,
  title = {Vortexability: A unifying criterion for ideal fractional Chern insulators},
  author = {Ledwith, Patrick J. and Vishwanath, Ashvin and Parker, Daniel E.},
  journal = {Phys. Rev. B},
  volume = {108},
  issue = {20},
  pages = {205144},
  numpages = {20},
  year = {2023},
  month = {Nov},
  publisher = {American Physical Society},
  doi = {10.1103/PhysRevB.108.205144},
  url = {https://link.aps.org/doi/10.1103/PhysRevB.108.205144}
}

@article{10.1063/1.5046122,
    author = {Haldane, F. D. M.},
    title = {The origin of holomorphic states in Landau levels from non-commutative geometry and a new formula for their overlaps on the torus},
    journal = {Journal of Mathematical Physics},
    volume = {59},
    number = {8},
    pages = {081901},
    year = {2018},
    month = {08},
    issn = {0022-2488},
    doi = {10.1063/1.5046122},
    url = {https://doi.org/10.1063/1.5046122},
}

@article{PhysRevB.102.165148,
  title = {Contrasting lattice geometry dependent versus independent quantities: Ramifications for Berry curvature, energy gaps, and dynamics},
  author = {Simon, Steven H. and Rudner, Mark S.},
  journal = {Phys. Rev. B},
  volume = {102},
  issue = {16},
  pages = {165148},
  numpages = {13},
  year = {2020},
  month = {Oct},
  publisher = {American Physical Society},
  doi = {10.1103/PhysRevB.102.165148},
  url = {https://link.aps.org/doi/10.1103/PhysRevB.102.165148}
}

@article{cai2023signatures,
  title={Signatures of fractional quantum anomalous Hall states in twisted MoTe2},
  author={Cai, Jiaqi and Anderson, Eric and Wang, Chong and Zhang, Xiaowei and Liu, Xiaoyu and Holtzmann, William and Zhang, Yinong and Fan, Fengren and Taniguchi, Takashi and Watanabe, Kenji and others},
  journal={Nature},
  volume={622},
  number={7981},
  pages={63--68},
  year={2023},
  publisher={Nature Publishing Group UK London},
  url = {https://www.nature.com/articles/s41586-023-06289-w}
}

@article{PhysRevLett.62.86,
  title = {Order Parameter and Ginzburg-Landau Theory for the Fractional Quantum Hall Effect},
  author = {Read, N.},
  journal = {Phys. Rev. Lett.},
  volume = {62},
  issue = {1},
  pages = {86--89},
  numpages = {0},
  year = {1989},
  month = {Jan},
  publisher = {American Physical Society},
  doi = {10.1103/PhysRevLett.62.86},
  url = {https://link.aps.org/doi/10.1103/PhysRevLett.62.86}
}

@article{PhysRevLett.62.82,
  title = {Effective-Field-Theory Model for the Fractional Quantum Hall Effect},
  author = {Zhang, S. C. and Hansson, T. H. and Kivelson, S.},
  journal = {Phys. Rev. Lett.},
  volume = {62},
  issue = {1},
  pages = {82--85},
  numpages = {0},
  year = {1989},
  month = {Jan},
  publisher = {American Physical Society},
  doi = {10.1103/PhysRevLett.62.82},
  url = {https://link.aps.org/doi/10.1103/PhysRevLett.62.82}
}

@article{zeng2023thermodynamic,
  title={Thermodynamic evidence of fractional Chern insulator in moir{\'e} MoTe2},
  author={Zeng, Yihang and Xia, Zhengchao and Kang, Kaifei and Zhu, Jiacheng and Kn{\"u}ppel, Patrick and Vaswani, Chirag and Watanabe, Kenji and Taniguchi, Takashi and Mak, Kin Fai and Shan, Jie},
  journal={Nature},
  volume={622},
  number={7981},
  pages={69--73},
  year={2023},
  publisher={Nature Publishing Group UK London},
  url={https://www.nature.com/articles/s41586-023-06452-3}
}

@article{park2023observation,
  title={Observation of fractionally quantized anomalous Hall effect},
  author={Park, Heonjoon and Cai, Jiaqi and Anderson, Eric and Zhang, Yinong and Zhu, Jiayi and Liu, Xiaoyu and Wang, Chong and Holtzmann, William and Hu, Chaowei and Liu, Zhaoyu and others},
  journal={Nature},
  volume={622},
  number={7981},
  pages={74--79},
  year={2023},
  publisher={Nature Publishing Group UK London},
  url={https://www.nature.com/articles/s41586-023-06536-0}
}

@article{PhysRevX.13.031037,
  title = {Observation of Integer and Fractional Quantum Anomalous Hall Effects in Twisted Bilayer ${\mathrm{MoTe}}_{2}$},
  author = {Xu, Fan and Sun, Zheng and Jia, Tongtong and Liu, Chang and Xu, Cheng and Li, Chushan and Gu, Yu and Watanabe, Kenji and Taniguchi, Takashi and Tong, Bingbing and Jia, Jinfeng and Shi, Zhiwen and Jiang, Shengwei and Zhang, Yang and Liu, Xiaoxue and Li, Tingxin},
  journal = {Phys. Rev. X},
  volume = {13},
  issue = {3},
  pages = {031037},
  numpages = {12},
  year = {2023},
  month = {Sep},
  publisher = {American Physical Society},
  doi = {10.1103/PhysRevX.13.031037},
  url = {https://link.aps.org/doi/10.1103/PhysRevX.13.031037}
}

@book{prange1987quantum,
  title={The quantum Hall effect},
  author={Prange, Richard E and Girvin, Steven M},
  year={1987},
  publisher={Springer}
}

@book{chakraborty2013quantum,
  title={The quantum Hall effects: integral and fractional},
  author={Chakraborty, Tapash and Pietil{\"a}inen, Pekka},
  volume={85},
  year={2013},
  publisher={Springer Science \& Business Media}
}

@book{jain2007composite,
  title={Composite fermions},
  author={Jain, Jainendra K},
  year={2007},
  publisher={Cambridge University Press}
}

@article{PhysRevLett.106.236804,
  title = {Fractional Quantum Hall States at Zero Magnetic Field},
  author = {Neupert, Titus and Santos, Luiz and Chamon, Claudio and Mudry, Christopher},
  journal = {Phys. Rev. Lett.},
  volume = {106},
  issue = {23},
  pages = {236804},
  numpages = {4},
  year = {2011},
  month = {Jun},
  publisher = {American Physical Society},
  doi = {10.1103/PhysRevLett.106.236804},
  url = {https://link.aps.org/doi/10.1103/PhysRevLett.106.236804}
}

@article{PhysRevLett.132.236503,
  title = {Microscopic Model for Fractional Quantum Hall Nematics},
  author = {Pu, Songyang and Balram, Ajit C. and Taylor, Joseph and Fradkin, Eduardo and Papi\ifmmode \acute{c}\else \'{c}\fi{}, Zlatko},
  journal = {Phys. Rev. Lett.},
  volume = {132},
  issue = {23},
  pages = {236503},
  numpages = {9},
  year = {2024},
  month = {Jun},
  publisher = {American Physical Society},
  doi = {10.1103/PhysRevLett.132.236503},
  url = {https://link.aps.org/doi/10.1103/PhysRevLett.132.236503}
}

@article{PhysRevB.92.035103,
  title = {Abelian and non-Abelian states in $\ensuremath{\nu}=2/3$ bilayer fractional quantum Hall systems},
  author = {Peterson, Michael R. and Wu, Yang-Le and Cheng, Meng and Barkeshli, Maissam and Wang, Zhenghan and Das Sarma, Sankar},
  journal = {Phys. Rev. B},
  volume = {92},
  issue = {3},
  pages = {035103},
  numpages = {7},
  year = {2015},
  month = {Jul},
  publisher = {American Physical Society},
  doi = {10.1103/PhysRevB.92.035103},
  url = {https://link.aps.org/doi/10.1103/PhysRevB.92.035103}
}

@article{PhysRevLett.85.5396,
  title = {Spontaneous Breakdown of Translational Symmetry in Quantum Hall Systems: Crystalline Order in High Landau Levels},
  author = {Haldane, F. D. M. and Rezayi, E. H. and Yang, Kun},
  journal = {Phys. Rev. Lett.},
  volume = {85},
  issue = {25},
  pages = {5396--5399},
  numpages = {0},
  year = {2000},
  month = {Dec},
  publisher = {American Physical Society},
  doi = {10.1103/PhysRevLett.85.5396},
  url = {https://link.aps.org/doi/10.1103/PhysRevLett.85.5396}
}

@article{PhysRevB.104.L121110,
  title = {Melting phase diagram of bubble phases in high Landau levels},
  author = {Villegas Rosales, K. A. and Singh, S. K. and Deng, H. and Chung, Y. J. and Pfeiffer, L. N. and West, K. W. and Baldwin, K. W. and Shayegan, M.},
  journal = {Phys. Rev. B},
  volume = {104},
  issue = {12},
  pages = {L121110},
  numpages = {5},
  year = {2021},
  month = {Sep},
  publisher = {American Physical Society},
  doi = {10.1103/PhysRevB.104.L121110},
  url = {https://link.aps.org/doi/10.1103/PhysRevB.104.L121110}
}

@article{He_2015,
doi = {10.1088/1367-2630/17/12/125005},
url = {https://doi.org/10.1088/1367-2630/17/12/125005},
year = {2015},
month = {dec},
publisher = {IOP Publishing},
volume = {17},
number = {12},
pages = {125005},
author = {He, Ai-Lei and Luo, Wei-Wei and Wang, Yi-Fei and Gong, Chang-De},
title = {Wave functions for fractional Chern insulators in disk geometry},
journal = {New Journal of Physics}
}

@misc{yu2025deconfinedquantumcriticalpoint,
      title={Deconfined Quantum Critical Point in Quantum Hall Bilayers}, 
      author={Guangyu Yu and Tao Xiang and Zheng Zhu},
      year={2025},
      eprint={2509.03079},
      archivePrefix={arXiv},
      primaryClass={cond-mat.mes-hall},
      url={https://arxiv.org/abs/2509.03079}, 
}
 
\end{document}